\documentclass[lettersize,journal]{IEEEtran}
\usepackage{booktabs}
\usepackage{xcolor}
\usepackage{amsmath,amsfonts}
\usepackage{algorithmic}
\usepackage{algorithm}
\usepackage{array}
\usepackage[caption=false,font=normalsize,labelfont=sf,textfont=sf]{subfig}
\usepackage{textcomp}
\usepackage{stfloats}
\usepackage{url}
\usepackage{verbatim}
\usepackage{graphicx}
\usepackage{cite}
\begin{document}

\title{Universal on-chip polarization handling with deep photonic networks }

\author{Aycan Deniz Vit$^{1}$, Ujal Rzayev$^{1}$, Bahrem Serhat Danis$^{1}$, Ali Najjar Amiri$^{2}$, Kazim Gorgulu$^{1}$, \\ Emir Salih Magden$^{1*}$%
\\
$^{1}$\IEEEmembership{Department of Electrical \& Electronics Engineering, Koç University, 34450, Istanbul, Turkey}\\
$^{2}$\IEEEmembership{Department of Electrical and Computer Engineering, Northwestern University, Evanston, IL 60208, USA}%
\thanks{*Corresponding author: esmagden@ku.edu.tr}}% <-this % stops a space

% The paper headers
% \markboth{Journal of \LaTeX\ Class Files,~Vol.~14, No.~8, August~2024}%
% {Shell \MakeLowercase{\textit{et al.}}: A Sample Article Using IEEEtran.cls for IEEE Journals}

% \IEEEpubid{0000--0000/00\$00.00~\copyright~2021 IEEE}
% Remember, if you use this you must call \IEEEpubidadjcol in the second
% column for its text to clear the IEEEpubid mark.

\maketitle

\begin{abstract}
We propose a novel design paradigm for arbitrarily
capable deep photonic networks of cascaded Mach-
Zehnder Interferometers (MZIs) for on-chip universal polarization handling. Using a device architecture made of cascaded Mach-Zehnder interferometers, we modify and train the phase difference between interferometer arms for both polarizations through wide operation bandwidths. Three proof-of-concept polarization handling devices are illustrated using a software-defined, physics-informed neural framework, to achieve user-specified target device responses as functions of polarization and wavelength. These devices include a polarization splitter, a polarization-independent power splitter, and an arbitrary polarization-dependent splitter to illustrate the capabilities of the design framework. The performance for all three devices is optimized using transfer matrix calculations; and their final responses are verified through 3D-FDTD simulations. All devices demonstrate state-of-the-art performance metrics with over 20 dB extinction, and flat-top transmission bands through bandwidths of 120 nm. In addition to the functional diversity enabled, the optimization for each device is completed in under a minute, highlighting the computational efficiency of the design paradigm presented. These results demonstrate the versatility of the deep photonic network design ecosystem in polarization management, unveiling promising prospects for advanced on-chip applications in optical communications, sensing, and computing.

\end{abstract}


\section{Introduction}

In recent years, guided-wave platforms, particularly silicon-on-insulator photonics, have experienced significant advancement with applications expanding from traditional communications and sensing to next-generation spectroscopy, optical computation, and artificial intelligence \cite{ref1a,ref2a,ref3a,ref4a,ref5a,ref6a,ref7a}. Managing varying input polarizations in these high-throughput applications remains a key requirement \cite{ref8a,ref9a,ref10a,ref11a}. Unintended polarization-dependent on-chip responses due to high-index contrast and waveguide aspect ratios lead to performance degradation including polarization-dependent losses and modified spectral characteristics. Since such limitations impair large-scale photonic systems, effective on-chip polarization management is essential for most applications \cite{ref12a}.

Previous studies have explored handling of different polarizations through a number of methods, yielding devices that are either polarization-independent or capable of selectively managing signals based on their  polarization state \cite{ref18a}, based on asymmetrical couplers \cite{ref13a,ref24a}, microring resonators \cite{ref14a}, bent directional couplers \cite{ref16a} and interferometers \cite{ref15a,ref21a}. Examples include polarization-independent power splitters using mode evolution \cite{ref20a}, polarization splitters and rotators with subwavelength grating claddings \cite{ref23a}, and polarization splitters achieved through adiabatic couplers \cite{ref25a}, multi-level tapers, or counter-tapered couplers \cite{ref17a,ref19a}. Polarization demultiplexing has also been demonstrated using integrated dielectric metasurfaces \cite{ref27a}. These application-specific prior examples are often limited in flexibility and bandwidth due to their reliance on specific waveguide geometries and wave coupling dynamics. Moreover, the diverse architectures and parameters in these designs highlight the lack of a universal framework for managing different input polarizations. Such a framework would ideally be adaptable to a wide range of functionalities and material platforms, enabling arbitrary polarization handling capabilities on-chip.

In this paper, we introduce a universal design paradigm for arbitrary polarization management using silicon-based deep photonic networks of cascaded Mach-Zehnder interferometers (MZIs), capable of handling both TE and TM polarizations over a wide range of wavelengths. Custom tapers with specific phase profiles in each MZI collectively form the network's transfer function based on the input polarization and wavelength. Using a physics-informed neural framework and gradient-based optimization, we iteratively train these networks for user-specified optical responses. We demonstrate this capability with three broadband devices: a polarization splitter, a polarization-independent power splitter, and a polarization-dependent arbitrary coupler. Designed in under a minute per device, these networks show state-of-the-art performance with flat-top transmission, 15–20 dB extinction ratios, $>120$ nm bandwidths, verified by 3D-FDTD simulations. Collectively, these devices provide a broad range of polarization management functionalities, highlighting the adaptability of our photonic network design within a unified system architecture. These capabilities can be extended to various material and fabrication platforms, positioning deep photonic networks as an effective solution for enhancing throughput in next-generation optical applications requiring versatile polarization control.

\section{Methodology}

\subsection{Deep Photonic Network Architecture}

\begin{figure*}[!t]
\centering
\includegraphics[width=7in]{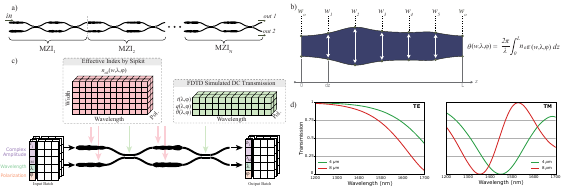}%
\label{fig1_label}
\caption{(a) A deep photonic network made up of n cascaded MZI blocks with 1-in 2-out port configuration is shown. (b) The phase delay section is one of the fundamental building blocks of an MZI that introduces a phase mismatch profile between interferometric arms. For both TE and TM polarizations, the effective index and the resulting phase delay as a function of wavelength is computed using the Silicon Photonics Toolkit library \cite{ref29a}. (c) An MZI block processes complex input amplitudes, wavelengths, and polarization in a batched manner by applying a linear transformation in which transmission coefficients and phase delay values are fetched from a database with auto-differentiation and fast data retrieval compatibility. (d) The first directional coupler is 4 $\mu$m-long; and the second directional coupler is 8 $\mu$m-long for every MZI in the network. These specific lengths are selected to ensure inputs at all wavelengths will experience non-zero coupling to the adjacent waveguide inside each MZI.
}
\end{figure*}

The deep photonic network architecture featuring cascaded MZIs has been previously demonstrated for optical power splitting and spectral duplexing \cite{ref28a}, and its general structure is illustrated in Fig 1a. For the specific architecture we built, the network follows a rectangular cascade of MZIs known as the “Clements” architecture \cite{ref28a,ref22a}, in order to maintain near-balanced optical path lengths and effectively reduce the network depth. Other topologies such as triangular or graph-like connected MZIs may also be chosen to implement similar optical functionality through adapting the simulation methods described. In this configuration, each individual MZI consists of the combination of two sets of directional couplers, and pairs of custom phase delay sections. By creating specific phase profiles between the two waveguide arms of the interferometer, these phase delays play a pivotal role in the specific routing of optical input as a function of wavelength and polarization. Each custom phase delay is a result of a specific selection of widths $w_1...w_n$, through which a continuous waveguide geometry is constructed as shown in Fig 1b. This construction enables waveguide tapers with unique phase profiles that can be configured for universal optical routing capabilities as a function of polarization.  For each custom taper, the phase delay $\theta$ is calculated from the effective indices of fundamental TE and TM polarizations through $\theta(w,\lambda,\varphi) = \frac{2\pi}{\lambda} \int n_{\mathrm{eff}}($w$, \lambda, \varphi) dz$.  Here, the effective index neff is extracted from the silicon photonics toolkit (SiPkit) \cite{ref29a}, an open-source software library allowing for rapid access to fundamental waveguide parameters between 1200-1700 nm of wavelength, for waveguide widths ranging from 400 to 480 nm. Importantly, the SiPkit library allows for the retrieval of these parameters in a manner that is compatible with automatic differentiation, making it possible to perform gradient-based optimizations of the waveguide widths in the custom tapers. The transmission response of directional couplers for fundamental TE and TM polarizations are obtained from 3D-FDTD simulations and are also recorded as a function of wavelength in the form of differentiable interpolations as shown in Fig 1d. In these networks, we use directional couplers with two different coupling lengths (4 $\mu$m and 8 $\mu$m) in each MZI to ensure that inputs at all wavelengths between 1200-1700 nm experience non-zero coupling to the adjacent waveguide. Different deep photonic networks either with exchanged positions of the two directional couplers or using entirely different couplers, can be optimized using similar techniques. Effective index data from SiPkit and 3D-FDTD simulated directional coupler transmissions are then used to compute the transfer matrix for each as  

\begin{equation}
\label{deqn_ex1a}
\begin{aligned}
T_i = & \begin{bmatrix}
      	  t_2(\lambda, \varphi)    & -jq_2(\lambda, \varphi) \\
      	  -jq_2(\lambda, \varphi) & t_2(\lambda, \varphi)    \\
           \end{bmatrix}
           \begin{bmatrix}
      	  \theta_{i3}  & 0            \\
      	  0 	  & \theta_{i4}   \\
           \end{bmatrix} \\
          & \begin{bmatrix}
      	  t_1(\lambda, \varphi)    & -jq_1(\lambda, \varphi) \\
      	  -jq_1(\lambda, \varphi) & t_1(\lambda, \varphi)    \\
           \end{bmatrix}
           \begin{bmatrix}
      	  \theta_{i1}  & 0            \\
      	  0 	  & \theta_{i2}   \\
           \end{bmatrix}	
\end{aligned}
\end{equation}

where $\lambda$ is wavelength, $\varphi$ is polarization, $\theta_{\mathrm{im}}$ is the corresponding phase at $m^{\mathrm{th}}$ phase delay in the $i^{\mathrm{th}}$ MZI, and $T_{\mathrm{i}}$ is the transfer matrix of the $i^\mathrm{th}$ MZI as labeled by $\mathrm{MZI}_{\mathrm{i}}$ in Fig 1a. Then, the overall response of the entire device is computed by propagating an optical input through each one of the transfer matrices of these constituent MZIs. This approach enables several computational advantages: To begin with, as the individual MZIs are independent building blocks of the photonic network, their transfer matrices can be constructed in a parallel manner, by cascading the responses of their constituent directional couplers and phase delays. On top of this parallelism, this architecture also enables batch processing techniques, allowing for simultaneous simulation of multiple inputs at different polarizations and wavelengths through the photonic network. Finally, each batch of optical inputs can be simulated through the network by using computational accelerators such as GPUs or TPUs \cite{ref30a}, making it possible to both retrieve the entire optical response with extreme computational efficiency and also iteratively train the geometrical device parameters to optimize the optical response for particular, application-specific functionalities including arbitrary polarization handling.

\subsection{Optimization Flow}

\begin{figure}[!t]
\centering
\includegraphics[width=3.5in]{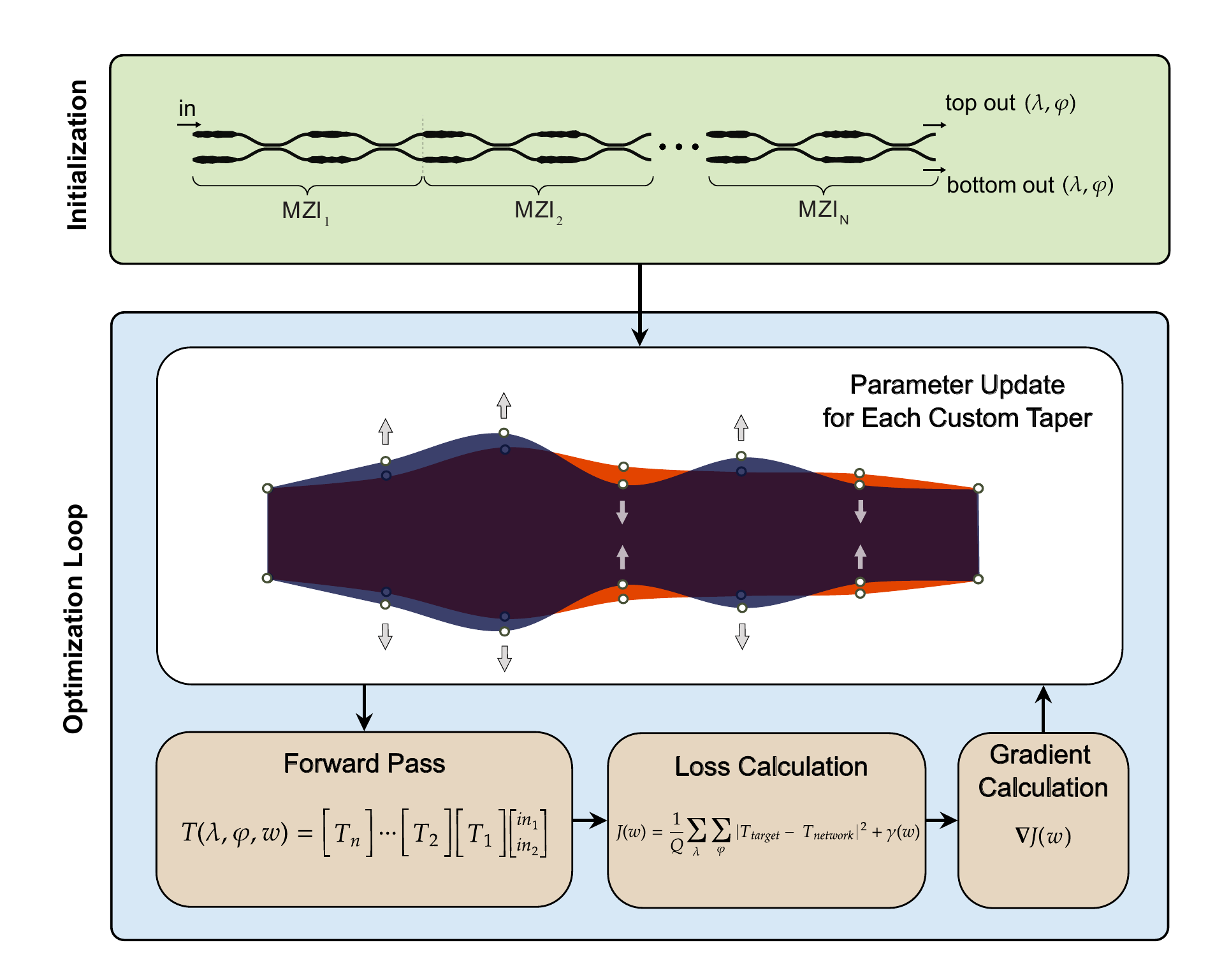}%
\label{fig2_label}
\caption{Illustration of the deep photonic network design paradigm including initialization and optimization stages.
}
\end{figure}

The optimization procedure for designing these deep photonic networks is illustrated in Fig. 2 with two main steps: device initialization and an optimization loop. First, the network architecture (input/output ports, numbers of MZI layers) and optical objectives are defined for TE/TM polarizations across multiple wavelengths. Then, a software-defined equivalent of this network is created using previously described automatic differentiation-compatible building blocks, architectural hyperparameters, and an end-to-end machine learning library \cite{ref31a}.

In the optimization loop, a loss function is defined as the difference between the target objective and the simulated transmission response at both polarizations, as shown in Eq. (2),  where $J$ is the loss function, $\lambda$ is wavelength, $\varphi$ is polarization, $T$ is power transmission, $\gamma$ is regularization factor as a function of widths of taper segments, and $Q$ is the number of wavelengths included in the target response.

\begin{equation}
\label{deqn_ex1a}
\begin{aligned}
J(w) = \frac{1}{Q} \sum_{\lambda} \sum_{\varphi} \left| T_{\mathrm{target}}(\lambda, \varphi) - T_{\mathrm{network}}(\lambda, \varphi) \right| ^2 + \gamma(w) 
\end{aligned}
\end{equation}

The gradient of this L2-loss is calculated via backpropagation, and used to iteratively update the trainable waveguide widths to minimize deviation from the target optical functionality. Optimal device geometry is achieved when the simulated response closely matches the target for both polarizations, per convergence criteria. During this optimization, several regularizations on waveguide widths are used to limit potential losses in custom tapers. To minimize potential scattering losses due to abrupt width transitions, the widths are constrained between 350 nm and 700 nm, around a standard width of 450 nm. Furthermore, several regularizations applied during the design process collectively reduce abrupt changes in waveguide and guide them toward the 450 nm default, ensuring a reliable and low-loss final device design. These regularizations yielded custom taper widths between 400 nm and 580 nm for our devices. Post-optimization, the physical layout is generated from its geometrical description for further simulation/verification and fabrication variation analysis.

This approach enables efficient and precise optimization of deep photonic networks for universal polarization handling, leveraging automatic differentiation and computational accelerators to rapidly explore and fine-tune device geometries. In the following section, we showcase the versatility of this method by demonstrating the design and performance of three distinct devices tailored for specific optical functionalities, including arbitrary polarization manipulation.

\section{Polarization-Handling Device Designs and Simulations}

Using our methodology, we demonstrate three 1-input, 2-output deep photonic networks for polarization handling. To showcase design versatility, we design three devices with increasingly complex polarization-handling tasks over a 120 nm bandwidth (1490-1610 nm). All three networks are constructed from five MZI layers, resulting in a total footprint of $4\times320 \mu m^2$ for each device. The first device, a polarization splitter, separates TE and TM inputs, directing each to its respective output port. The second, a polarization-independent 50/50 power splitter, evenly divides the input power between the two outputs regardless of polarization. The final device equally splits TE-polarized input power to two outputs, while directing all of the TM-polarized input to only one of the output ports. For each device the transmission characteristics were verified with 3D-FDTD simulations; and fabrication tolerance analyses were performed.

\subsection{Polarization Splitter}

\begin{figure*}[!t]
\centering
\includegraphics[width=7in]{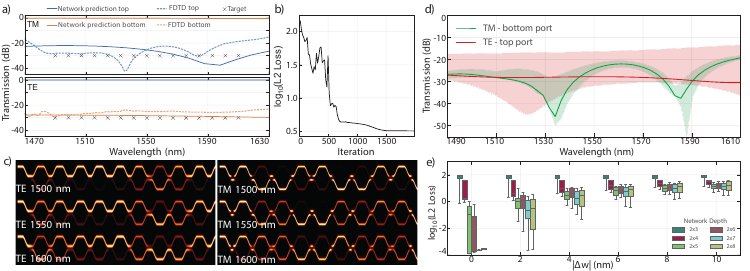}%
\label{fig3_label}
\caption{(a) Polarization splitter deep photonic network achieves an extinction ratio exceeding 20 dB for both polarizations and an operating bandwidth of 120 nm around 1550 nm. 3D-FDTD response of the device is also shown with dashed curves for the two polarizations. (b) Evolution of error (total L2-loss) during optimization. (c) Electric field intensity for TE and TM polarizations at wavelengths of 1500, 1550 and 1600 nm. Light is routed to the respective output ports at both wavelengths depending on the input polarization. (d) Change in transmission under fabrication imperfections, simulated for waveguide width variations of up to 10 nm. Possible maximum transmission deviations in the response are shown with the shaded regions.  (e) L2-loss as a function of waveguide over/under etch offset up to $\pm$ 10 nm, for network depths from 3 to 8 cascaded MZIs.}
\end{figure*}

In this first demonstration, a photonic network was created and optimized for the purpose of separating TE and TM polarized inputs into two separate outputs. The transmission response results are plotted in Fig 3a. Here, the solid curves indicate the expected response from the optimization of the network; and the dashed curves represent 3D-FDTD simulation results of the same devices. As illustrated by these results, TE/TM polarization splitting functionality was achieved with an extinction ratio of over 20 dB for both polarizations, and the insertion loss primarily originating from directional couplers remains below 0.45 dB. The match between the 3D-FDTD results and the transfer-matrix response verifies the physical accuracy of the network simulations, which are computationally much more efficient than fully vectorial electromagnetic simulations of comparable accuracy. The convergence of the loss function shown in Fig 3b indicates a successful optimization process in 58.1 seconds, thanks to the highly parallelizable architecture of the software-defined model described above. These findings illustrate that while these demonstrated networks yield performance metrics comparable to other polarization splitters in literature \cite{ref32a,ref33a}, their optimization and design can be performed through much faster algorithmic procedures with 3D-FDTD-level physical accuracy, and design flexibility that extends beyond simple polarization splitting tasks.

The electric field intensity obtained from the device’s 3D-FDTD response is plotted in Fig 3c at 1500, 1550 and 1600 nm, for both TE and TM polarizations. These field profiles reveal that even though both inputs are routed to the specified output port of the optimized network, the number of crossings between the two arms of the cascaded interferometers and the overall interferometric pattern can be drastically different throughout the broadband operation spectrum. This is a unique advantage of the design framework described, as the resulting optimized photonic networks are able to satisfy a large number of objectives at multiple different wavelengths and polarizations systematically.

Next, we calculate the tolerance of the network to fabrication-induced variations and plot the transmission responses under waveguide width variations of up to $\pm$ 10 nm with shaded regions in Fig 3d. The nominal transmissions under ideal fabrication conditions are also plotted with the solid curves for reference. The results indicate that the extinction ratio remains over 20 dB and it is consistent with measurement results of other polarization splitters demonstrated previously \cite{ref32a,ref33a}. The results also show that the TM polarization response exhibits better fabrication tolerance compared to the TE polarization response, attributed to the slower rate of change in the effective index of TM polarization, which reduces its phase delay sensitivity to waveguide width changes. 

\begin{figure*}[!t]
\centering
\includegraphics[width=7in]{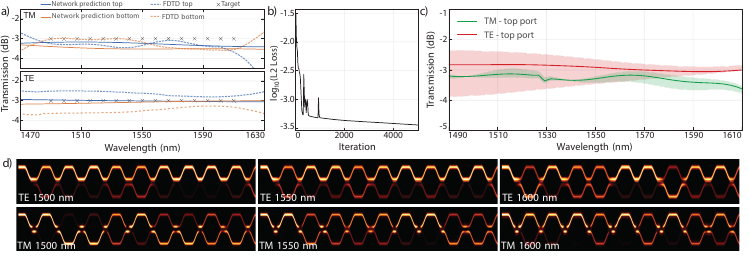}%
\label{fig4_label}
\caption{(a) Polarization-independent power splitter transmission results from transfer matrix calculations (solid) and 3D-FDTD simulations (dashed). The plotted response indicates a 1 dB bandwidth as wide as 120 nm for both TE and TM polarizations. (b) Evolution of error (total L2-loss) during optimization. (c) Change in transmissions under fabrication imperfections for width variations of up to 10 nm. Possible maximum transmission deviations in the response are shown with shaded regions. (d) Electric field intensity recorded from 3D-FDTD simulations for TE and TM polarizations at wavelengths 1500, 1550 and 1600 nm. Light is evenly split between the two outputs at both polarizations.
}
\end{figure*}

The overall performance of the device and its fabrication tolerance are a function of the number of layers, as this parameter directly influences the number of trainable widths, the network capability, and the device footprint. In Fig. 3e, we plot optimization results for devices ranging from 3 layers to 8 layers, where a total of 25 devices have been optimized for each network depth. In the ideal case when no fabrication errors are considered ($\Delta w=0$), the results indicate that the network performance drastically improves with increasing network depth, up to a depth of 5 layers. The vertical spreads illustrate the statistical variability in the final performance for the 25 networks with 4, 5, and 6 layers, highlighting the importance of network initialization \cite{ref26a}. Deeper networks with 7 and 8 layers also demonstrate ideal performance with much less statistical variability among the 25 devices optimized. This can be attributed to the increased capability of the optical network, making it easier for the optimizer to arrive at optimal solutions regardless of the initialization state. We also note that the L2-loss experiences a slight increase from $10^{-4}$ towards $3 \times 10^{-4}$ with deeper networks. This is a direct result of the insertion loss experienced through the directional couplers. This loss is modeled by our computation depicted in Fig. 1, accumulates through successive layers of MZIs, and is reflected in the reduced performance of the network.

For the same devices of each network depth, we then analyze fabrication tolerance by specifying a nonzero $\Delta w$. As anticipated, all devices experience performance deterioration with increasing fabrication errors, indicated by the general increase in L2-loss with growing $| \Delta w |$. However, longer devices experience larger degradation in their performance, due to the accumulation of phase errors through longer waveguides. This behavior points to a design tradeoff between overall network capability and expected fabrication tolerance, as deeper networks provide greater optical capability at the expense of larger insertion loss and worse fabrication tolerance. In general, the selection of the network depth hyperparameter requires careful consideration of these factors for the specific fabrication platform chosen, and is a critical part of the design stage. For our devices, we find that a network depth of 5 layers is sufficient for the desired overall device performance while also yielding acceptable fabrication tolerance. For the other two classes of devices we demonstrate in the following sections, we find that while the specific analysis metrics slightly differ, the general behavior described here remains the same. This phenomenon is also quantitatively explained by the individual errors corresponding to the TE and TM polarizations plotted in Fig 3e, where the TE-polarized input experiences larger differences from the target transmission for all given waveguide width variations.

\subsection{Polarization-independent Power Splitter}

The next polarization-specific device we demonstrate is a dual-polarization 50/50 power splitter, for which the details are given in Fig 4. The transmission results from the transfer matrix calculations (solid) and 3D-FDTD simulations (dashed) are shown in Fig 4a for both polarizations. Throughout the entire optimized bandwidth, the resulting transmissions show agreement with the target specifications, demonstrating uniform power-splitting capabilities for both TE and TM polarizations with transmissions at both output ports remaining within $\pm$0.6 dB of the -3 dB target. The entire optimization procedure is completed in 54.9 seconds as illustrated by the gradual progression of the loss shown in Fig 4b. For this specific example, the optimization may also be terminated after 2000 iterations by specifying convergence criteria based on the relative improvement of loss between iterations. The 3D-FDTD results in Fig 4d illustrate the operation of the device at two different wavelengths on either side of the operation bandwidth, for both TE and TM polarizations. As before, the optimizer specifically configures the custom tapers for achieving 50/50 output transmissions for all target wavelengths specified by systematically adjusting the phase difference in each one of the constituent MZI arms.

We analyze the fabrication tolerance of the polarization-independent splitter in Fig 4c. The transmission response for the two polarizations are plotted with red and green shaded areas, for width variations of up to 10 nm in Fig 4c. These results indicate that the expected transmissions may change by a few percent for TM-polarized inputs, and up to 5-10\% for TE-polarized inputs, when compared to the device with no width variations (solid curves).

\subsection{Polarization-dependent Arbitrary Power Splitter}

\begin{figure*}[!t]
\centering
\includegraphics[width=7in]{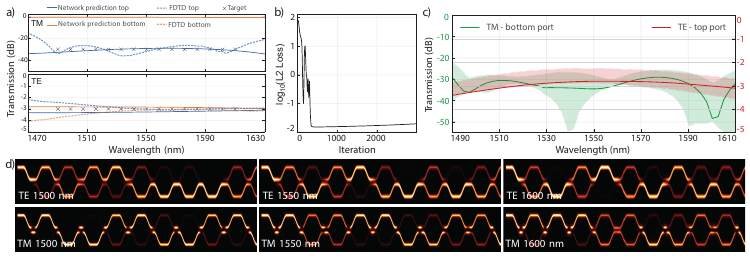}%
\label{fig5_label}
\caption{(a) Transmission results for a proof-of-concept, arbitrary polarization-handling device. TM-polarized input is directed to the top output port with an extinction ratio of over 20 dB; and the TE-polarized output is split evenly between the two output ports. Transfer matrix (solid curves) and 3D-FDTD responses (dashed curves) are plotted. (b) Evolution of error during optimization with convergence to the specified optical target functionality. (c) Change in transmission under fabrication imperfections with width variations of up to 10 nm, shown by the shaded regions. (d) Electric field intensity for TE and TM-polarized inputs at wavelengths 1500, 1550 and 1600 nm.
}
\end{figure*}

Finally, we detail the design and operation of a more complex device with arbitrary polarization-dependent transmission characteristics in Fig 5. In this device, the objective is specified such that the TE-polarized input is split evenly between the two output ports (50/50), while the entire TM-polarized input is routed to the top output port (100/0). This ability to arbitrarily specify the desired transfer functions for TE and TM polarized inputs demonstrates functional capabilities that extend beyond traditional approaches. While more straightforward polarization splitter or polarization-independent power splitter examples have been demonstrated as referenced previously, the deep photonic network platform is the first approach that systematically allows for such arbitrary polarization handling tasks to be efficiently realized, to the best of our knowledge. The transmission characteristics for this optimized device are plotted in Fig 5a with transfer matrix results (solid curves) and 3D-FDTD simulations (dashed curves). The TM mode achieves better than 20 dB extinction; and the TE mode undergoes less than 0.3 dB deviation from the -3 dB target across the entire 120 nm of operation bandwidth, as separately verified by the FDTD simulations. The insertion loss of the device, primarily resulting from accumulated losses in the waveguide bends of the directional couplers, remained below 0.48 dB, pointing to the overall device efficiency and platform capability. The optimization procedure is completed in 48.2 seconds as plotted in Fig 5b, highlighting the computational efficiency of the demonstrated design paradigm. The 3D-FDTD results in Fig 5d illustrate the broadband operation of the device, and also the scalability of the automated design paradigm demonstrated to arbitrary polarization handling tasks. In Fig 5c, the transmission under waveguide width variations of up to 10 nm is shown with the shaded regions. Once again, the TM polarization exhibits better fabrication tolerance. 

For more complex objectives, our design procedure allows for more cascaded layers with increasing design flexibility. While five layers were sufficient for all devices here, deeper networks can be easily explored for performance and fabrication tolerance. The networks also generalize to more than two outputs, with specifiable transfer functions for arbitrary input-output pairs. The simulation results from the three devices we study here underscore our networks' capabilities in achieving arbitrarily specified polarization handling with high extinction ratios and minimal insertion losses. We also note that the same optimization procedure may also be used to mitigate fabrication-induced errors by simultaneously optimizing over/under-etched versions of the same device \cite{ref28a}. Moreover, the demonstrated methodology is not restricted to silicon and can be extended to other material platforms, depending on system requirements. By offering a framework that can be standardized and applied across different materials or fabrication platforms, this design paradigm enables universal polarization handling capabilities within a unified device architecture. These capabilities highlight deep photonic networks as a scalable, universally applicable architecture for specific polarization handling in advanced optical communications, sensing, and computing.

As detailed in Table~\ref{tab:comparison_bench}, our deep photonic network devices demonstrate competitive performance compared to state-of-the-art designs ~\cite{ref18a, ref24a, ref32a, ref33a, ref37a, ref38a, ref20a, ref34a, ref35a, ref36a}. Our polarization splitters achieve an extinction ratio exceeding 20 dB with less than 0.45 dB excess loss over a 120 nm bandwidth. The polarization-independent power splitter maintains a splitting imbalance below 1 dB and an excess loss under 0.5 dB across the same spectrum. Beyond their strong performance, our devices are also designed in under one minute each, demonstrating several orders of magnitude better computational efficiency than the designs for comparable structures. This efficiency offers a major advantage over conventional approaches, which require prior knowledge of mode coupling dynamics, and over other inverse design methods that rely on iterative electromagnetic simulations. The rapid optimization and flexibility of our method enable the design of complex, general-purpose polarization handling devices that are otherwise challenging to realize efficiently.

\begin{table*}[htbp]
\caption{Performance Comparison of On-Chip Polarization Handling Devices}
\label{tab:comparison_bench}
\centering
\scriptsize
\begin{tabular}{@{}l l l l l l l l l@{}}
\toprule
\textbf{Reference} & \textbf{Structure} & \textbf{Design Timeframe} & \textbf{Capability} & \textbf{Footprint [µm$^2$]} & \textbf{ER [dB]} & \textbf{Split Imb. [dB]} & \textbf{EL [dB]} & \textbf{Bandwidth [nm]} \\
\midrule
\multicolumn{9}{@{}l}{\textit{Rotating/Non-Rotating Polarization Splitters}} \\
\cite{ref18a} & Double-etched Taper & traditional design (hours) & function-specific & $\sim 788$ & $>20$ & --- & $<0.8$ & $<120$ \\
\cite{ref33a} & Double-etched DC & traditional design (hours) & function-specific & $\sim 95$ & $>25$ & --- & $<0.5$ & 30 \\
\cite{ref24a} & Taper + ADC & traditional design (hours) & function-specific & $\sim 1250$ & $>10$ & --- & $<1.5$ & $<40$ \\
\cite{ref32a} & Multi-mode WG + ADC & traditional design (hours) & function-specific & $237.5$ & $>20$ & --- & $<0.6$ & $<85$ \\
\cite{ref37a} & Inverse Design & 12 hr & general purpose & $2$ & $>15$ & --- & $<0.9$ & $<40$ \\
\cite{ref38a} & Inverse Design & 19 hr & general purpose & $51$ & $>15$ & --- & $<1.0$ & $<50$ \\
Our Work & Deep Photonic Network & 58.1 sec & general purpose & 1280 & $>20$ & --- & $<0.5$ & 120 \\
\midrule
\multicolumn{9}{@{}l}{\textit{Polarization-Independent Power Splitters}} \\
\cite{ref20a} & Slotted WG + Tapers & traditional design (hours) & function-specific & $\sim 500$ & --- & $<1.00$ & $<1.0$ & $<390$ \\
\cite{ref34a} & Modified MMI Junction & traditional design (hours) & function-specific & $\sim 2.3$ &  --- & $<1.21$ & $<0.3$ & $>80$ \\
\cite{ref35a} & SWGs & traditional design (hours) & function-specific & $\sim 32$ & --- & $<0.02$ & $<0.8$ & $<170$ \\
\cite{ref36a} & Cascaded Bent DCs & traditional design (hours) & function-specific & $\sim 600$ & --- & $<2.00$ & $<1.0$ & $<110$ \\
Our Work & Deep Photonic Network & 54.9 sec & general purpose & 1280 & --- & $< 1.00$ & $<0.5$ & 120 \\
\bottomrule

\multicolumn{9}{@{}p{\dimexpr\linewidth-2\tabcolsep}@{}}{%
\vspace{0.5ex}%
\scriptsize% 
\textbf{ER:} Extinction Ratio. \textbf{Split Imb. (Splitting Imbalance):} Power difference between the different output ports relative to input power. \textbf{EL:} Excess Loss. \textbf{Bandwidth:} 0.5 dB bandwidth, defined as the wavelength range where transmissions remain within 0.5 dB of their optimal values. \textbf{ADC:} Asymmetric Directional Coupler. \textbf{MMI:} Multi-Mode Interference. \textbf{SWG:} Subwavelength Grating. \textbf{WG:} Waveguide.
}

\end{tabular}
\end{table*}

\section{Conclusion} 
In summary, we presented a pioneering deep photonic network methodology for designing devices for arbitrary on-chip polarization manipulation. Using custom tapers and a software-defined neural network framework, we designed broadband devices with state-of-the-art metrics including flat-top transmission, high extinction ratios, and wide operational bandwidths. Our numerical results specifically show >20 dB extinction ratios and >120 nm operational bandwidths (centered at 1550 nm) for various polarization-handling tasks. This design framework's demonstrated scalability and adaptability can advance integrated photonics, with potential applications in optical communication, sensing, and computing.

\section*{Acknowledgments}
This work was supported by the Scientific and Technological Research Council of Turkey (TUBITAK) under grant number 119E195, and the Marie Sklodowska-Curie Fellowship (no 101032147) through the Horizon 2020 program of the European Commission.

%\section{References Section}
%You can use a bibliography generated by BibTeX as a .bbl file.
% BibTeX documentation can be easily obtained at:
% http://mirror.ctan.org/biblio/bibtex/contrib/doc/
% The IEEEtran BibTeX style support page is:
% http://www.michaelshell.org/tex/ieeetran/bibtex/

\end{document}